\documentclass[final,10pt,3p,times]{elsarticle}
\usepackage{amssymb}
\usepackage{amsmath}

\usepackage{natbib}
\usepackage{graphicx} 
\usepackage{braket, slashed} 
\renewcommand{\eqref}[1]{(\ref{#1})} 
\usepackage{appendix}

\usepackage{todonotes} 

\newcommand{\edit}[2]{#2}  

\newcommand{\logexp}{\mathop{\mathrm{logexp}}}

\DeclareMathOperator{\KL}{S_{SJ}}

\DeclareMathOperator{\Diag}{Diag}
\def\rr{{\mathbb R}}

\DeclareMathOperator*{\argmin}{arg\,min}
\usepackage{soul}
\usepackage[normalem]{ulem} 

\begin{document}

\title{Dual formulation of the maximum entropy method applied to analytic continuation of quantum Monte Carlo data}

\author[add1,add2]{Thomas Chuna}
\author[add3]{Nicholas Barnfield}%
\author[add1,add2]{Tobias Dornheim}
\author[add4]{Michael P. Friedlander}
\author[add5]{Tim Hoheisel}

\address[add1]{Center for Advanced Systems Understanding (CASUS), D-02826 Goerlitz, Germany}
\address[add2]{Helmholtz-Zentrum Dresden-Rossendorf (HZDR), D-01328 Dresden, Germany}
\address[add3]{Department of Statistics, Harvard University, Cambridge, Massachusetts 02138, USA}
\address[add4]{Department of Computer Science and Mathematics, University of British Columbia, Vancouver, BC V6T 1Z4, Canada}
\address[add5]{Department of Mathematics and Statistics, McGill University, Montreal, Quebec H3A 0G4, Canada}

\date{\today}

\begin{abstract}
Many fields of physics use quantum Monte Carlo techniques, but struggle to estimate dynamic spectra via the analytic continuation of imaginary-time quantum Monte Carlo data. One of the most ubiquitous approaches to analytic continuation is the maximum entropy method (MEM). We supply a dual Newton optimization algorithm to be used within the MEM and provide analytic bounds for the algorithm's error. The MEM is typically used with Bryan's controversial algorithm [Rothkopf, "Bryan’s Maximum Entropy Method" Data 5.3 (2020)]. We present new theoretical issues that are not yet in the literature. Our algorithm has all the theoretical benefits of Bryan's algorithm without these theoretical issues. We compare the MEM with Bryan's optimization to the MEM with our dual Newton optimization on test problems from lattice quantum chromodynamics and plasma physics. These comparisons show that in the presence of noise the dual Newton algorithm produces better estimates and error bars; this indicates the limits of Bryan's algorithm's applicability. We use the MEM to investigate authentic quantum Monte Carlo data for the uniform electron gas at warm dense matter conditions and further substantiate the roton-type feature in the dispersion relation.
\end{abstract}

\maketitle

\section{Introduction}
Path integral Monte Carlo (PIMC) simulations are used to estimate imaginary-time correlation functions (ITCF) $F(\tau)$, which are one- and two-particle Green functions. The ITCFs are associated with dynamical spectra $S(\omega)$, which are useful because they predict measurable real-time quantities. However, to extract spectra from ITCFs an analytic continuation to real time is required. This involves inverting the integral transform
\begin{align}
    F(\tau) =  \frac{1}{2 \pi} \int_{-\infty}^{+\infty} A(\tau, \omega)  S(\omega)\, d\omega,
\end{align}
where $\tau \in [0,\beta)$. Across different applications, computational budgets prohibit us from collecting sufficient data that would-- even in the ideal situation-- be sufficient to uniquely invert the transformation. In addition to this difficulty, the kernel takes one of the forms $A(\tau, \omega) = \exp(- \omega \tau)$ or $A(\tau, \omega) = \frac{\exp(- \omega \tau)}{1 \pm e^{-\beta \omega} }$. Thus, the analytic continuation problem is, or simplifies to, inverting the Laplace transform, which is an exponentially ill-posed problem, \textit{i.e.}, the singular values of the transformation matrix decay to zero exponentially fast~\cite{epstein2008badtruth}. Therefore, the inversion problem needs to be regularized in order to make the inversion well-posed. In this paper, we propose a dual Newton algorithm for solving an entropy-regularized linear least-squares problem that alleviates the difficulties associated with the analytic continuation problem. The dual Newton algorithm we describe efficiently scales to large problems and is accompanied by an error analysis that characterizes its accuracy from observable quantities.  


The maximum entropy method (MEM), as proposed by Bryan~\cite{bryan1990algorithm} and subsequently applied to the analytic continuation problem by Gubernatis et al.~\cite{gubernatis1991MEM}, typically serves as the benchmark for new methods~\cite{kim2018comparison, tripolt2019comparison, yang2023combining, Fischer2018smoothBR, fuchs2010analytic}. These comparisons are complicated by the fact that Bryan's maximum-entropy approach uses an optimization algorithm that does not solve the original entropy-regularized least-squares formulation, but only an approximation~\cite{bryan1990algorithm, rothkopf2020bryan}. Bryan's algorithm relies on a singular value decomposition (SVD) of the discrete kernel $A \in \rr^{N_\tau \times N_\omega}$ and represents the entire kernel $A = U \Sigma V^T$ by $N_s$ of right singular vectors in $V$, denoted
\begin{align}\label{eq:BryanSubstitution}
    A \rightarrow \tilde{V}^T,
\end{align}
where $N_s < N_\tau$. This substitution eliminates the left singular vectors $U$ and singular values $\Sigma$. These comparisons are also complicated by the fact that the relative entropy, which Bryan's algorithm optimizes, includes a term driving the elements of the spectral function to sum to the same value as the elements of the default model (\textit{i.e.}, Bayesian prior). In this work, we will show that although this assumption is valid for zero-temperature systems, the MEM is often applied outside this limit. Our dual Newton algorithm addresses both of these issues.

Our comparisons between the MEM with Bryan's algorithm and the MEM with a dual Newton algorithm reflect the theoretical advantages of our algorithm. Our dual Newton optimizer produces better solutions and uncertainty estimates. With respect to its solutions, the MEM with the dual Newton algorithm uses the transformation kernel's entire basis so the dual Newton's solution differs from Bryan's solution. In particular, the additional basis vectors allow the dual Newton algorithm to estimate functional forms that are problematic for Bryan's algorithm. With respect to the uncertainty, noise causes the posterior weighting function for MEM with Bryan's algorithm to favor more regularized solutions than the MEM with the dual Newton algorithm. The implication is that the variance of proposed solutions diminishes and causes the MEM with Bryan's algorithm to artificially suppress uncertainty estimates. Error estimates are of utmost importance for scientific applications where the true solution is not known. To further address this uncertainty issue, we provide analytic expressions of the error bounds for a solution obtained via the dual Newton.

The paper is partitioned as follows. Section~\ref{sec_priorwork} describes prior work. Section~\ref{sec_methods} formulates the optimization problem, describes the dual approach, and presents the analytic error bounds. Section~\ref{sec_results} presents numerical comparisons between the dual Newton and Bryan's algorithm. We conduct three investigations. First, we consider Asakawa et al.'s zero-temperature $\rho$-meson benchmark, which is relavant to the field of lattice quantum chromodynamics~\cite{Asakawa2001}. Next, we introduce a new finite temperature benchmark, using the completed Mermin collisional uniform electron gas dynamic structure factor (DSF)~\cite{chuna2024conservative}. This example is relavant to the field of high energy density physics. Finally, we apply the method to authentic path integral Monte Carlo data of the uniform electron gas at warm dense matter conditions. Section~\ref{sec_conclusions} discusses the conclusions we draw from our numerical results. \ref{app:newton-cg} gives an overview of our dual optimization algorithm.

\section{Prior Work}\label{sec_priorwork}

Many methods have been applied to the analytic continuation problem across different fields of physics. The Backus-Gilbert (BG) method~\cite{backus1968, backus1970} and its smooth variant~\cite{hansen2019extraction} has been used to estimate spectral densities of quarkonia. The Bayesian reconstruction method (BRM)~\cite{rothkopf2013improved, Burnier2013Bayesian} and its smoothed variant~\cite{kim2015lattice, Fischer2018smoothBR} have similarly been applied. Genetic algorithms have been used to construct spectral functions of ultracold $^4$He~\cite{Boninsegni1996, ferre2016dynamic, Kora_PRB_2018} and $^3$He~\cite{Dornheim_SciRep_2022}. Artificial neural networks have been used to estimate band structures of the Hubbard model~\cite{Yoon_PRB_2018,Fournier_PRL_2020}. Stochastic analytic inference (SAI) methods have been used to estimate the spectral function of a BCS super conductor~\cite{beach2004identifying}, band structures of the Hubbard model~\cite{fuchs2010analytic}, and spectral functions in LQCD~\cite{shu2015stochastic}. The maximum entropy method (MEM) has been widely applied to estimate quantities of interest in LQCD~\cite{Asakawa2001}, Hubbard model~\cite{fuchs2010analytic}, Dynamical Mean Field Theory~\cite{Kotliar_RMP_2006}, and many more. Additionally, specialized methods have been developed for specific applications, including the Nevanlinna representation of spectral functions with noise-free data~\cite{Nevanlinna_PRL_2021}, the Hamburger problem formulation based on the frequency moments of a given spectral density~\cite{tkachenko_book}, and methods using additional constraints derived from known physical properties of dynamic spectra~\cite{dornheim2018stochasticsamplingalg, Groth2019}. Comparative studies suggest that there is no established ``best" approach to the analytic continuation problem~\cite{kim2018comparison, tripolt2019comparison, yang2023combining, Fischer2018smoothBR, fuchs2010analytic}.

We discuss the more general algorithms, which take any given analytic continuation kernel and any given ITCF data. Axiomatic arguments conclude that minimizing a least-squares fidelity objective function with relative-entropy regularization does not introduce erroneous correlations into the solution~\cite{shore1980Axiomatic1, shore1980Axiomatic2, skilling1988axioms}, as illustrated by examples outlined by Jarrel and Gubernatis~\cite{jarrell1996MEM}. Therefore, entropic regularization, as seen in MEM, BRM, and SAI methods, is central to our approach. Across comparisons of these three methods, the MEM estimate tends to be the smoothest of these entropic estimates. Comparing MEM and BRM, Kim et al.~\cite{kim2015lattice} show that MEM estimates can be produced by the BRM if a smoothness regularization is added to the BRM objective function. A primary motivation for adding the smoothness regularization was that the BRM estimates often contained spurious wiggles not present in the MEM estimate~\cite{Fischer2018smoothBR}. Comparing MEM and SAI, Fuchs et al.~\cite{fuchs2010analytic} conclude that the SAI method produces spectral functions that are less regularized and consequently show more pronounced features than the MEM, but their SAI method is orders of magnitude more computationally expensive than the MEM.

Both Kim et al.~\cite{kim2015lattice} and Fuchs et al.~\cite{fuchs2010analytic} use a MEM based on Bryan's optimization algorithm. Bryan argues that his controversial substitution \eqref{eq:BryanSubstitution} is justified when the gradient of the fidelity term lies in the singular space, aiming to simplify the problem from an ill-conditioned $N_\omega$-dimensional optimization to a well-conditioned $N_s$-dimensional optimization. This reduction in dimensionality and improved conditioning both decrease the computational cost. Essentially, Bryan’s approach removes singular vectors associated with small singular values, thereby eliminating flat search directions from the optimization process. However, regularizing through singular values enhances a solution's smoothness, limiting MEM's ability to resolve pronounced features---an observation consistent with the findings of Kim et al. and Fuchs et al. Furthermore, a recent study has identified specific example functions that cannot be adequately represented by this truncated basis set~\cite{rothkopf2020bryan}. This highlights a need for a maximum entropy algorithm that retains the advantages of Bryan's method (\textit{i.e.}, reduced search space and improved conditioning) without excluding singular vectors.

In this work, we introduce an optimization algorithm designed to replace Bryan's optimization algorithm. While Bryan's algorithm uses a Levenberg-Marquardt optimizer to solve a simplified version of the primal problem~\cite{bryan1990algorithm}, our approach uses a Newton-Krylov optimizer to solve the \emph{dual problem}. Although the dual formulation of entropic regularization is well established~\cite{rioux2020maximum, marechal1997unification, marechal1998principle, le1999new}, it has not previously been applied to the analytic continuation problem. Both the primal and dual formulations lead to the same unique solution; however, optimizing the dual formulation offers certain advantages, including guaranteed differentiability and reduced dimensionality. In particular, the dual problem is formulated within the column space of the kernel rather than its row space. For a discretized transformation kernel $A \in \rr^{N_\tau \times N_\omega}$, this reduces the optimization dimension from $N_\omega$ to $N_\tau$. For instance, in the work of Asakawa et al.~\cite{Asakawa2001}, the dual formulation would reduce the optimization dimension from $n=600$ to $m \leq 30$ without any loss of information. Additionally, differentiability allows for the use of second-order optimizers, which outperform the Levenberg-Marquardt optimizer. In summary, the dual approach enables the application of a higher-order optimizer to a better-conditioned problem within a reduced search space, while preserving all basis vectors.

\section{Maximum Entropy Method \label{sec_methods}}
In this section we formulate analytic continuation, formulate dual optimization, and formulate error bounds for the dual solution. We do not discuss the MEM procedure in full detail. There are many well-organized works where the MEM is outlined \cite{jarrell1996MEM, Asakawa2001, bryan1990algorithm}. The method assumes there exists a collection of solutions which optimize the least-squares fidelity term plus entropic regularizer at different regularizer weights $\alpha$. The core of the MEM is Gull's derivation of the Bayesian posterior, which indicates how to properly combine these solutions \cite{gull1989MEMBayesianWeighting}. Importantly, the MEM procedure does not specify how the solutions are to be obtained. Typically Bryan's algorithm, which uses a modified Levenberg-Marquardt algorithm, is used to obtain these solutions \cite{bryan1990algorithm}. Yet, for any given $\alpha$ the MEM's cost function has a unique minimum, which can be discussed independent of the optimization algorithm (\textit{e.g.}, Levenberg-Marquardt, ADAM, Newton, conjugate gradient) or the approach (\textit{i.e.}, primal or dual) used to obtain it. In this work, we leave Gull's Bayesian analysis intact and replace Bryan's primal modified Levenberg-Marquardt algorithm with a better performing dual Newton optimization algorithm. In practice, we leave our MEM code intact and a call different optimizer subroutine.

\subsection{Noisy Linear Model of PIMC Data \label{subsec_model}}
In the most general form, PIMC methods generate samples of ITCFs $F(\textbf{x}, \tau\mid\, \beta )$, where $\textbf{x}$ is the spatial coordinate, $\tau$ is imaginary-time, and $\beta$ is the inverse temperature. By definition, the desired spectra are the Fourier transform of the correlation functions. Fourier transforming in the spatial dimension is trivial and we arrive at $F(k, \tau)$ where $k$ is the wavenumber. However, Fourier transforming the temporal component $\tau$ is non-trivial. Most PIMC simulations are done in imaginary time $\tau = - i t$. Thus the Fourier kernel $\exp(i \omega t)$ becomes a Laplace kernel, and the Laplace transform of the data is \footnote{Since $S(k, \omega)$ contains poles, solving \eqref{eq:acproblem} is an analytic continuation problem \cite{Ichimaru1991}. }
\begin{equation} \label{eq:acproblem}
    F(k, \tau) \equiv  \frac{1}{2 \pi} \int_{-\infty}^{+\infty} \exp(- \omega \tau)  S(k, \omega) d \omega + \eta(k,\tau).
\end{equation}
The factor of $2 \pi$ converts $\omega$ from $\text{rad}/s$ to $1/s$ and the sampling noise $\eta(k, \tau)$ arises from the PIMC algorithm (\textit{e.g.} Metropolis-Hastings).
This work will focus on inverting \eqref{eq:acproblem}, though other works continue on to produce the fermion [boson] kernels $A(\tau, \omega) = \frac{\exp(- \omega \tau)}{1 \pm e^{-\beta \omega}}$.

For finite temperature simulations, periodic boundary conditions introduce the temperature, \textit{i.e.}, $F(k,\tau) = F(k,\beta - \tau)$ and $T = 1/\beta$. This leads most research efforts to consider the inversion of periodic Laplace transforms (\textit{e.g.}, Hansen et al. \cite{hansen2019extraction} or Ferre et al. \cite{ferre2016dynamic}), expressed as,
\begin{equation} \label{eq:acproblem-periodic}
    F(k, \tau) \equiv  \frac{1}{2 \pi}\int_0^\infty \left( \exp(- \omega \tau)  +  \exp(- \omega (\beta - \tau)) \right) S(k, \omega) d \omega + \eta.
\end{equation}
We have suppressed the arguments of the noise $\eta$ for compactness. This equation is derived from \eqref{eq:acproblem} using the detailed balance condition $S(k,-\omega) = e^{-\beta \omega} S(k,\omega)$~\cite{quantum_theory}. 

The Laplace \eqref{eq:acproblem} or periodic Laplace \eqref{eq:acproblem-periodic} both have strengths and weaknesses. The strength of the non-periodic approach \eqref{eq:acproblem} is that the normalization of $S(k,\omega)$, given by
\begin{align}
    F(k,\tau=0) = \frac{1}{2 \pi} \int_{-\infty}^\infty S(k, \omega) d \omega,
\end{align}
is known and can be used to scale the correlation function's $\tau=0$ entry to $1$, which normalizes $S(k,\omega)$.
The weakness of the Laplace transform is that the exponential suppression arising from the detail balance relation means that hundreds of numerical zeros are being fit in the $\omega < 0$ domain.
By comparison, the strength of the periodic approach is that the $\omega < 0$ degrees of freedom are eliminated. 
The weakness of the periodic kernel is that the positive normalization of $S(k,\omega)$
\begin{align}\label{eq:positive-normalization}
    Z = \frac{1}{2 \pi} \int_0^\infty S(k, \omega) d \omega,
\end{align}
is not known. To see that the positive normalization is not known, consider \eqref{eq:acproblem-periodic} at $\tau=0$:
\begin{align}
    F(k,\tau=0) =  \frac{1}{2 \pi} \int_0^\infty \left( 1 +  \exp(- \omega \beta) \right) S(k, \omega) d \omega.
\end{align}
When the temperature is zero, the second term can be neglected, but when the temperature is finite the frequency dependent term persists and the value of $F(k,\tau=0)$ does not match the positive normalization \ref{eq:positive-normalization}. A positive normalization $Z$ must either be supplied or searched for. The implications of not knowing the value of $Z$ is described in the problem formulation subsection \ref{subsec_ProblemFormulation}. In this work, we opt for the periodic Laplace transform; this choice matches with the majority of the literature we referenced in section~\ref{sec_priorwork}. 

Another practical consideration is the discretization of the transform. PIMC simulations have discrete time variables. We denote the discrete $\tau$ with index $i$ as $\tau_i = i \Delta \tau$ where $i \in \{0,\ldots,N_\tau-1\}$; both $\Delta \tau$ and $N_\tau$ are set by the PIMC simulation. Computational capacity limits PIMC simulations to a discretization of size $N_\tau \approx 10^2-10^3$. Additionally, we impose the discretization of the frequency $\omega$ as $\omega_j = j \Delta \omega$ where $\Delta \omega$ is the resolution and $j \in \{1,\ldots,N_\omega\}$. The resolution needs to be sufficient to resolve the relevant features in $S(k,\omega)$. 
Typically $N_\omega \approx 10^3-10^4$ is sufficient.
Additionally, PIMC simulation have periodic space variable, which leads to a discrete wavenumber $k$.
For a given $k$ value, the discrete formulation of either \eqref{eq:acproblem} or \eqref{eq:acproblem-periodic} is given by
\begin{align}\label{eq:acproblem-discrete}
    b_i = \Delta \omega \sum_j A_{i,j} x_j + \eta_i.
\end{align}
We dropped the $k$ index for compactness and we have expressed the discrete kernel $A(\tau_i,\omega_j)$ as matrix $A_{i,j}$, the discrete spectra $S(k, \omega_j)$ as the vector $x$,  the discrete noise $\eta(k,\tau_i)$ as the vector $\eta_i$, and the discrete correlation function $F(k,\tau_i)$ as the vector $b_i$. 

\subsection{Problem formulation \label{subsec_ProblemFormulation}}

Equation~\eqref{eq:acproblem-discrete} presents a numerically tractable formulation of the inversion problem. However, when $N_\tau < N_\omega$ the inversion of $A$ becomes underdetermined, meaning the solution is not unique. Additionally, the discretized Laplace transform produces a matrix $A_{i,j}$ whose singular values decay exponentially fast, often approaching numerical zero. This rapid decay leads to severe ill-conditioning, making a direct solution via least squares both unstable and unreliable.

In such cases, regularization techniques are essential to constrain the problem and obtain a unique, stable solution. Regularization helps to mitigate the ill-posedness by incorporating prior information, ensuring the solution reflects both the data and the typical smoothness of a dynamic structure factor (DSF).


\subsubsection{KL-regularized least-squares}

The MEM minimizes a relative entropy term alongside a generalized least-squares fidelity term. The optimization problem is formulated as 
\begin{equation}\label{eq:primal}
    \min_{x\in\rr^{N_\omega}_+} \, \alpha \KL(x\mid \mu) + \frac{1}{2} \|Ax - b\|^2_{C^{-1}},
\end{equation}
where $\|z\|^2_{C^{-1}}=z^T C^{-1}z$ denotes the scaled 2-norm, $C$ represents the covariance matrix of the error in the observation vector $b$, and regularization parameter $\alpha \geq 0$ is a positive regularization parameter that balances the importance of fitting the observed measurements versus matching the Bayesian prior $\mu$. The term
\begin{align}\label{eq:S_sj}
 \KL(x\mid\mu) = \sum_{j=1}^n \mu_j - x_j + x_j \log(x_j/\mu_j)
\end{align}
is the Shannon--Jaynes (SJ) relative entropy\footnote{Relative entropy (Kullback--Leibler Divergence) belongs to the family of $f$-divergences, which are often interpreted as pseudo-metrics for probability measures.} between discrete distributions $x$ and $\mu$. Going forward we refer to \eqref{eq:S_sj} as the relative entropy. We adopt the standard convention that $0\log0=0$. Note that where $\mu_i=0$ then $x_i$ must also be zero to render $\KL(x\mid\mu)$ finite. The relative entropy encodes the conventional relative entropy term in additional to a linear term
\begin{equation}\label{eq:soft_constraint}
 \sum_{j=1}^n \mu_j - x_j,
\end{equation}
which enforces a soft-constraint on the positive normalization of $x$ with that of $\mu$. As an option, our inversion code allows the user to omit the soft constraint in \eqref{eq:soft_constraint}, optimizing instead the conventional relative entropy. \edit{}{In this work, we include the soft constraint to make direct comparisons.} 

\subsubsection{Normalization}

As discussed in connection with the periodic formulation~\eqref{eq:acproblem-periodic}, the normalization $Z$ defined in~\eqref{eq:positive-normalization} is not known. An additional source of uncertainty in the normalization accrues from the quantization errors in the discretization~\eqref{eq:acproblem-discrete}. The prior $\mu$ will have its own normalization $Z_\mu=\sum_i\mu_i$, but as shown by the functional ~\eqref{eq:smooth-map-x}, the discrete normalized distribution with maximum-entropy takes the form
\begin{equation}\label{eq:boltzmann}
 \frac{x}{Z} = \frac{\mu_j\exp(\sum_{i=1}^{N_\omega} A_{i,j} y_i)}{\sum_{\hat j=1}^{N_\tau}\mu_{\hat j}\exp(\sum_{i=1}^{N_\omega} A_{i,\hat j} y_i)},
\end{equation}
where the scalars $y_i$ are Lagrange multipliers corresponding to the dual maximum-entropy problem; see~\cite[Section~3]{kapurEntropyOptimizationPrinciples1992a} and~\eqref{eq:smooth-map-x}.
The form of \eqref{eq:boltzmann} is to be expected and matches the form of the grand cannonical Boltzmann factor.
Note that because the numerator and denominator of this expression are both linear in the prior $\mu$, we can see that scaling prior $\mu_j$ by $Z_\mu$ has no bearing on the maximum-entropy solution $x$.

For any fixed normalization $Z$, the maximum-entropy solution has negative entropy given by the value function
\begin{align} \label{eq:scaledprimal}
    V(Z) = \min_{x \in Z \Delta_{N_\omega}}  p(x, Z \mid \mu) 
\end{align}
where
\begin{align} \label{eq:S_sj-scaled}
    p(x,Z\mid\mu):= Z \KL'\left( \left. \frac{x}{Z} \right| \frac{\mu}{Z} \right) + \sum_{j=1}^n \mu_j - x_j +  \frac{1}{2 \alpha} \|Ax - b\|^2_{C^{-1}}
\end{align}
and 
\begin{align}\label{eq:KL}
 \KL'(x\mid\mu) = 
 \begin{cases}
      \sum_{j=1}^n x_j \log(x_j/\mu_j), & \text{if } x \in \Delta_{N_\omega}\\
      \infty,              & \text{otherwise,}
    \end{cases}
\end{align}
is the modified relative entropy term from~\eqref{eq:S_sj}. The identity $Z \KL'\left( \left. \frac{x}{Z} \right| \frac{\mu}{Z} \right) = \KL'(x\mid \mu)$ is applied in~\eqref{eq:scaledprimal} because solutions not normalized to $1$ incur an infinite penalty.

We thus aim to find the normalization $Z$ that solves the problem
\begin{align} \label{eq:scaledprimal_opt}
    \min_{Z > 0} V(Z).
\end{align}
This is equivalent to the problem~\eqref{eq:primal} in terms of achieving the same optimal value. 

\subsubsection{Dual of KL-regularized least squares}
The objective function given in~\eqref{eq:scaledprimal}, referred to as the \emph{primal} problem, is nonsmooth, necessitating the use of constrained or nonsmooth optimization algorithms. Such algorithms, however,  are typically less efficient and reliable than those designed for smooth unconstrained optimization~\cite[Section~17.2]{NoceWrig06}. Given the strict convexity of the KL divergence in~\eqref{eq:KL}, we can leverage the duality between strict convexity and differentiability~\cite[Theorem 11.13]{rockafellar_VariationalAnalysis_2009}.
\edit{Neglecting $\sum_{j=1}^n \mu_j - x_j$ from~\eqref{eq:scaledprimal}, the}{The} dual of the KL-regularized least-squares problem~\eqref{eq:scaledprimal} is the minimization, over $y\in\rr^{N_\tau}$, of the strongly convex and smooth objective function
\begin{equation}\label{eq:dual}
    \min_{y\in\rr^{N_\tau}}
    \, d(y, Z\mid \mu),
\end{equation}
where
\begin{align}
    d(y, Z\mid \mu) :=\frac{\alpha}{2} y^T C y - \langle b,y \rangle + Z \logexp (A^* y - c \mid \mu ) - Z \log(Z),    
\end{align}
is the dual objective function, and the log-sum-exp function
\begin{equation}\label{eq:logexp}
    \logexp(z\mid\mu) \equiv \log\sum_{j=1}^n\mu_j\exp(z_j)
\end{equation}
arises from the Legendre-Fenchel transform of the KL divergence, as detailed in Rockafellar and Wets~\cite[Example 11.2]{rockafellar_VariationalAnalysis_2009}.

A straightforward approach to optimizing~\eqref{eq:scaledprimal} involves two subproblems: first, for a given $Z$, optimize $p(x, Z \mid \mu)$, then search over $Z$ for the unique optimum. In this way, \eqref{eq:scaledprimal} corresponds to~\eqref{eq:primal}, where the feasible set of solutions is restricted to the $Z$-simplex. 

A one-to-one correspondence between the primal and dual solutions is established by Fenchel-Rockafeller duality~\cite[Theorem 11.39 and Example 11.41]{rockafellar_VariationalAnalysis_2009}. This relationship is defined by the smooth mappings
\begin{subequations}\label{eq:primal-dual-maps}
\begin{align}
  \label{eq:smooth-map-x}
  x(y) &= Z \, \nabla\logexp( A^* y \mid \mu), \\
  \label{eq:smooth-map-y}
  y(x) &= \alpha^{-1} C^{-1}( b-Ax ),
\end{align}
where
\begin{align}
\nabla\logexp(z\mid\mu) \equiv \frac{\mu\odot\exp(z)}{\mu^T\exp(z)}.
\end{align}
\end{subequations}
Here, $\exp(z)$ denotes the Hadamard (elementwise) product, so $\mu \odot \exp(z)$ represents a vector with components $\mu_j\exp(z_j)$, and $\exp(z)$ denotes a vector whose components are $\exp(z_j)$.



In summary, while the primal problem \eqref{eq:primal} is nonsmooth and thus requires gradient descent optimization, our associated dual problem \eqref{eq:dual} is smooth and can be optimized with using second-order solvers (\textit{e.g.}, Newton's Method). Furthermore, the primal problem is formulated in a $N_\omega$-dimensional column space and the dual problem is formulated in the lower $N_\tau$-dimensional row space. These benefits come with no drawbacks as the primal solution can be easily recovered from the dual solution via \eqref{eq:smooth-map-x},

In~\ref{app:newton-cg} we describe a Newton-Krylov method that leverages the particular structure of the objective function and searches over positive normalizations $Z$. However, a variety of other algorithms are also suitable since the dual problem~\eqref{eq:dual} is smooth and amenable to standard smooth optimization techniques.

\subsection{Perturbation Analysis}\label{sec:pertubation}



In this section, we examine how the computed primal solution is affected by various sources of error. Specifically, we analyze three key sources: first, we assess how perturbations in the dual solution influence the corresponding primal solution. Second, we analyze how perturbations in the vector of observations $b$ influence the primal solution. Finally, we investigate how perturbations in the unknown regularization parameter $\alpha$ influence the primal solution. Understanding these relationships is crucial for assessing the robustness and reliability of the solutions obtained through our approach. In all cases, sensitivity analysis is performed under a fixed $Z$ setting.

\subsubsection{Relationship between dual error and primal error}

For brevity in notation, we set $A_C := C^{-1/2}A$ and $b_C := C^{-1/2}b$ in this section. Let $y_p$ be the exact optimal solution of the dual problem~\eqref{eq:dual} for a fixed set of parameters $p=(b,\alpha)\in\rr^{N_\tau}\times \rr_{++}$, and let $x_p:=x(y_p)$ be the corresponding primal solution obtained via the map~\eqref{eq:smooth-map-x}. Since $x(y)$ is a composition of the adjoint of the linear map $A_C$ with $Z \logexp$, which has a $Z$-Lipschitz gradient, it follows that $x(y)$ is Lipschitz continuous with Lipschitz constant $Z \| A_C\|$ \cite[Example~5.15]{firstorderBeck}. For any parameterizations $p=(b,\alpha)$ and $p'=(b',\alpha')$, we obtain the upper bound
\begin{equation}\label{eq:error-x}
  \|x_p - x_{p'}\|\le Z \| A_C\|\cdot\|y_p-y_{p'}\|,
\end{equation}
where $\| \cdot \|$ denotes the 2-norm, and hence $\| A_C \|$ is the largest singular value of $A_C$. Equation~\eqref{eq:error-x} shows that any error in the inferred primal solution is bounded by a constant multiple of the error in the dual solution. 
An immediate application of this bound concerns the sequence of dual iterates $y^k$ generated by the Newton-Krylov method described in~\ref{app:newton-cg}. Specifically, the error in corresponding primal iterate $x(y^k)$ is bounded by the error in $y^k$ scaled by the norm of the operator $A_C$:
\begin{equation}\label{eq:error-xk}
\|x(y^k)-x^*\|\le Z \|A_C\|\cdot\|y^k-y^*\|,
\end{equation}
where $(x^*,y^*)$ represents the exact primal-dual solution. This bound establishes the validity of the dual approach: as the dual iterates converge to the dual optimum, the corresponding primal iterates also converge to the optimum at the same rate.

\subsubsection{Relationship between data perturbations and primal solution}

We examine how errors or changes in the observation data $b$ affect the solution of our primal problem. To do so, we consider the \emph{solution map} of the dual problem, which links the input data $b$ and regularization parameter $\alpha$ to the unique optimal solution.

The dual problem can be reformulated by completing the square in the dual objective function $d$, which yields
\begin{align*}
P(b_C/\alpha)
    &:= \argmin_{y\in\rr^m}\left\{ \frac{\alpha}2\|y-(1/\alpha)  b_C\|^2 + Z \logexp( A_C^*y)
     \right\}\\
    &= \argmin_{y\in\rr^m}\,d(y).
\end{align*}
Observe that the solution map $P$ is the \emph{proximal map} corresponding to the $\logexp$ function \eqref{eq:logexp} ~\cite[Chapter~6]{firstorderBeck}. Under mild conditions-- which are met by $f$-- the proximal map is nonexpansive~\cite[Proposition~23.8]{bauschke_ConvexAnalysisMonotone_2017}. This means that for any two vectors $b$ and $b'$,
\begin{align}
    \|y_b - y_{b'}\|\equiv\lVert P(b_C / \alpha) - P(b_C' / \alpha) \rVert  \leq \frac{1}{\alpha} \rVert b_C - b_C' \rVert.
\end{align}
Combining this bound with~\eqref{eq:error-x}, we deduce
\begin{equation}\label{eq:perturbation-in-b}
\lVert x_b - x_{b'} \rVert  \leq (Z/\alpha) \cdot  \lVert  C^{-1} \rVert\cdot \lVert A \rVert \cdot \rVert b - b' \rVert
\end{equation}
for all $b, b' \in \rr^m$. In other words, the optimal-solution map~\eqref{eq:smooth-map-x} for the primal problem is $(Z \alpha^{-1} \lVert C^{-1} \rVert \cdot \lVert A \rVert)$-Lipschitz with respect to the observation vector $b$.

This bound implies that if $b'$ represents the noiseless data and $b$ represents noisy data, then for a given $\alpha$, the solution $x$ obtained from noisy data will differ from the noiseless solution $x'$ by no more than the right hand side of \eqref{eq:perturbation-in-b}. As $\alpha \rightarrow \infty$, the solution becomes independent of the data $b$, illustrating the potential drawback of choosing highly regularized solutions.

\subsubsection{Relationship between regularization parameter and primal solution}

We analyze the sensitivity of the solution map to changes in the regularization parameter $\alpha$. The solution map associate the input parameters, such as $\alpha$, with the unique optimal solution to the problem. Here, we examine how variations in $\alpha$ affect the solution.

Consider the dual solution $y_\alpha$ for a fixed regularization parameter $\alpha$, assuming $A$ and $b$ are fixed. The dual problem can be reformulated by defining the function $F: \rr^{1} \times \rr^m \to \rr^m$ as
\begin{align}
    F(\alpha, y) = \alpha y - b_C + Z A_C D_z \logexp(z\mid\mu)|_{z=A^Ty},
\end{align}
where the partial gradient with respect to $y$ is
\begin{align}
    \nabla_y F(\alpha, y) = \alpha I + Z A_CD(y)A_C^T,
\end{align}
with
\begin{equation}\label{eq:logexp-Hessian}
D(y) = \nabla^2_z \logexp(z\mid\mu)|_{z=A^*_Cy} = Z \left( \Diag(x(y)) - x(y)x(y)^T \right),
\end{equation}
and $x(y)$ defined by~\eqref{eq:smooth-map-x}.
Since $F(\alpha, y_\alpha) = 0$ and $\nabla_y F(\alpha, y)$ is nonsingular for all $\alpha > 0$, we can apply the implicit function theorem~\cite[Theorem 1B.1]{DontRock} to obtain 
\begin{equation}
\begin{aligned}
y'_\alpha &= - \nabla_y F(\alpha, y_\alpha)^{-1} \nabla_{\alpha} F(\alpha, y_\alpha)
\\ &= - (\alpha I + Z A_C D(y_\alpha) A_C^T)^{-1}y_\alpha,
\end{aligned}
\end{equation}
where $y'_\alpha$ is the derivative of the solution map $\alpha \mapsto y_\alpha \in \argmin_y d(y\mid\mu)$. Because the corresponding primal solution $x(y_\alpha)$ lies in the probability simplex $\Delta_{N_\omega}$, which is compact, we have $\lVert y_\alpha \rVert \leq \alpha^{-1}(\lVert b_C \rVert + \lvert A_C \rVert)$ by \eqref{eq:smooth-map-y}. With $\lVert (\alpha I + Z A_C S(y_\alpha) A_C^T)^{-1} \rVert \leq \alpha^{-1}$, we derive
\[
\lVert y'_\alpha \rVert \leq \frac{\lVert A_C \rVert + \lVert b_C \rVert}{\alpha^2}.
\]
Written as a local perturbation bound, we obtain
\[
\lVert y_\alpha - y_{\alpha'}\lVert \leq \left(\frac{\lVert A_C \rVert + \lVert b_C \rVert}{\min(\alpha'^2, \alpha^2)}\right) \cdot\mid\alpha - \alpha'\mid.
\]
Combining this expression with~\eqref{eq:error-x}, yields
\begin{align} \label{eq:error-alpha}
\lVert x_\alpha - x_{\alpha'}\lVert \leq Z \cdot \lVert C^{-1} \rVert^2 \cdot \lVert A \rVert \cdot \left(\frac{\lVert A \rVert + \lVert b \rVert}{\min(\alpha'^2, \alpha^2)}\right) \cdot\mid\alpha - \alpha'\mid
\end{align}
for $\alpha'$ near $\alpha$. This bound shows that as $\alpha, \alpha' \to 0$, the bound weakens significantly, indicating greater sensitivity to small values of $\alpha$.

We note that uniform bounds (i.e., globally Lipschitz) are known; however, obtaining them requires more advanced concepts from convex analysis, as discussed by Bonnans and Shapiro~\cite[\S{4}]{Bonnans2000}. In essence, these bounds differ only by a factor of $2$, meaning that, up to this factor, \eqref{eq:error-alpha} quantifies the impact of different regularization choices $\alpha$ and $\alpha'$ on the solution.

\section{Results \label{sec_results}}
In this section, we consider analytic continuation problems relevant to the quark-gluon plasma (QGP) and warm dense matter. We demonstrate that the MEM can be used with the dual Newton optimization algorithm and we compare against the ubiquitous implementation of the MEM with Bryan's algorithm. The key goals are assessing whether sharp peaks, arising from a spectral function's poles, can be captured and whether spurious peaks arise in relatively smooth regions. In short, we focus on uncertainty quantification.

\subsection{LQCD Test Problems}
We consider the extraction of quarkonia spectral functions from lattice quantum chromodynamics (LQCD) PIMC simulations. The peaks in a spectral function represent bound states for a quark/anti-quark pair and as temperature increases these peaks broaden. At some critical temperature the peaks smooth away, indicating that the quark and anti-quark cannot form a bound state. In application, knowing the melting temperature of various quarkonia allows experimental facilities like the Relativistic Heavy Ion Collider or the Large Hadron Collider to estimate the temperature of the quark gluon plasmas they create. For more details see one of the excellent reviews of recent progress using PIMC on QGP \cite{petreczky2012QGPreview, bazavov2015QGPreview, datta2015QGPreview, rothkopf2020QGPreview}.

\subsubsection{Synthetic correlation samples from zero-temperature $\rho$ meson}
For this LQCD benchmark, we consider Asakawa, Nakahara, and Hatsuda's seminal work which introduced the MEM with Bryan's algorithm to quarkonia spectral function \cite{Asakawa2001}. In particular, we study the synthetic correlation data generated from the ``realistic spectral function'' of a charged $\rho$ meson. The charged $\rho$ meson spectral function is parameterized by the relativistic Breit-Wigner function and the relativistic Breit-Wigner's parameters were fit to reproduce the cross section for $e^{+} e^{-}$ annihilating into hadrons \cite{Shuryak1993}. Spectral functions are by definition $S(k=0,\omega)$ and often denoted as $\rho(\omega)$ in the LQCD community. We consider the spectral function $\rho(\omega) = \omega^2 x(\omega)$, where $x$ is given
\begin{equation}\label{eq:rhomesonSPF}
    x(\omega) = \frac{2}{\pi} \left( F_\rho^2 \frac{\Gamma_\rho(\omega) m_\rho}{(\omega^2 - m_\rho^2)^2 + \gamma_\rho^2 m_\rho^2 } + \frac{1}{8\pi}(1+\frac{\alpha_s}{\pi})\frac{1}{1+e^{(\omega_0-\omega)/\delta}}\right).
\end{equation}
The pole residue informs $F_\rho = m_\rho /g_{\rho \pi \pi}$ and 
\begin{align}
    \Gamma_\rho (\omega) = \frac{g_{\rho \pi \pi}^2}{48 \pi} m_\rho \left(1 - \frac{4 m_\pi^2}{\omega^2} \right)^{3/2} \theta(\omega - 2 m_\pi).
\end{align}
The empirical values of the constants are given by $m_\rho = 0.77$ GeV, $m_\pi = 0.14$ GeV, $g_{\rho \pi \pi} = 5.45$, $\omega_0=1.3$ GeV, $\delta = .02$ GeV. We multiply $x(\omega)$ with kernel $A_{i,j} = e^{\tau_i \omega_j} \omega_j^2$ to produce our noiseless synthetic correlation data $b$. We discretize using $\omega_j = j \Delta \omega$ where $j \in \{1,\ldots,N_\omega\}$, $N_\omega=600$, and $\Delta \omega =  0.01$ GeV as well as $\tau_i = i \Delta \tau$ where $i \in \{0,\ldots,N_\tau-1\}$, $N_\tau=30$, and $\Delta \tau = 0.431 \, \text{GeV}^{-1}$. We introduce noise $\eta(\tau_i) \sim \mathcal{N}(0, \sigma_i^2 )$ to the clean signal to construct synthetic samples. Following Asakawa et al., the variance is given by
\begin{align}\label{eq:syntheticnoise}
    \sigma_i^2 = \sigma^2 \frac{\tau_i + \epsilon}{\Delta \tau} b_i
\end{align} 
where $\sigma^2$ is a noise level we will vary and the perturbation $\epsilon=10^{-6}$ is introduced to ensure $\sigma_{i=0} > 0$. Since the noise is proportional to $\tau$ the signal is more precise at small $\tau$, which helps the method to resolve features at large $\omega$. Using this noise we generate $N_s=1000$ samples and estimate the data $b$ and the ecovariance matrix $C_{i,j} = \delta_{i,j} \sigma_i^2 / \sqrt{N_s}$ from those samples. This differs slightly from Asakawa et al. who generated a single sample and used the true covariance matrix; our method is more reflective of what is done in practice. We normalize the data $b_i$ and standard deviation $\sigma_i$ by the first entry of the data $b_0$. Anecdotally, this makes $\sum_i x_i < 1$. We use a flat prior $\mu = 0.257$ as our default model and normalize it by $b_0$. We then compute a collection of solutions $x_\alpha(\omega)$ over a domain of $\alpha$'s. From this larger $\alpha$ domain, we select a subdomain [$\alpha_\textrm{min}, \alpha_\textrm{max}$] where the posterior distribution-- derived by Gull \cite{gull1989MEMBayesianWeighting}-- $P(\alpha\mid b,\mu)$ and the maximum of that distribution $P(\alpha^*\mid b,\mu)$ are related by the bound $P(\alpha\mid b,\mu) \leq 0.1 \times P(\alpha^*\mid b,\mu)$. We compute the estimate $\hat{x}$ from the collection of solutions as 
\begin{subequations}
\label{eq:Bayesianweighting}
\begin{align}
    \hat{x}_j = \frac{\sum_\alpha x_j(\alpha) \, P(\alpha \mid b,\mu)}{ \sum_\alpha  P(\alpha\mid b,\mu)},
\end{align}
where the index $\alpha$ is the $\alpha$ values within [$\alpha_\textrm{min}, \alpha_\textrm{max}$] and $x_\alpha$ is the corresponding solution. The posterior weighting function is 
\begin{align}\label{eq:posterior}
    P(\alpha \mid b, \mu) \propto \exp \left( \alpha \KL(x \mid \mu)  - L  (x \mid b) + \frac{1}{2} \sum_k \log \frac{\alpha}{\alpha + \lambda_k} \right).
\end{align}
\end{subequations}
Here $\lambda_k$ is the $k^\mathrm{th}$ eigenvalue of the real symmetric matrix $\Lambda_{j,j'}(\alpha) = \sqrt{x_j(\alpha)} \frac{\partial^2 L}{\partial x_j(\alpha) \partial x_{j'}(\alpha)} \sqrt{x_{j'}(\alpha)}$.
In the typical fashion, we estimate the error as $\pm 2 \sqrt{ \widehat{(x^2)} - (\hat{x})^2 }$.

\begin{figure}
    \centering
    \includegraphics[width=.33\textwidth]{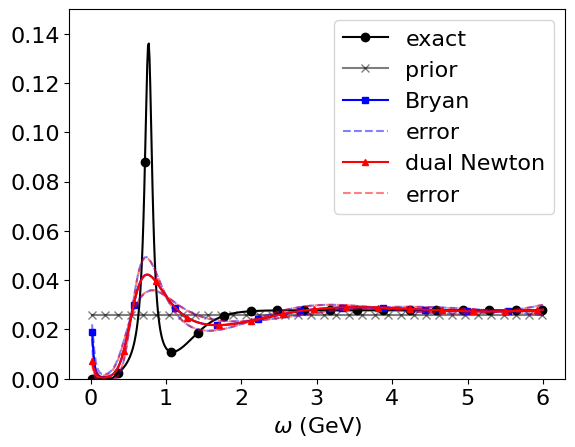}%
    \includegraphics[width=.33\textwidth]{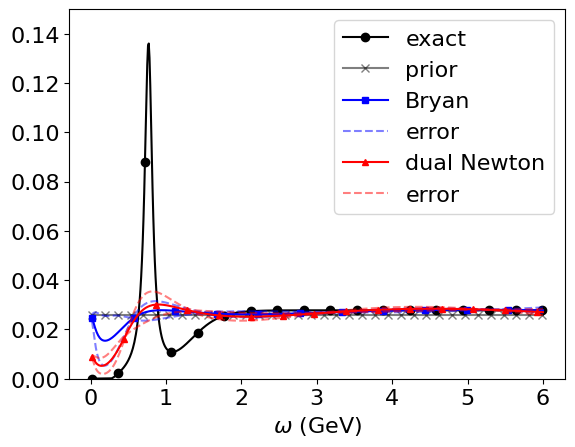}%
    \includegraphics[width=.33\textwidth]{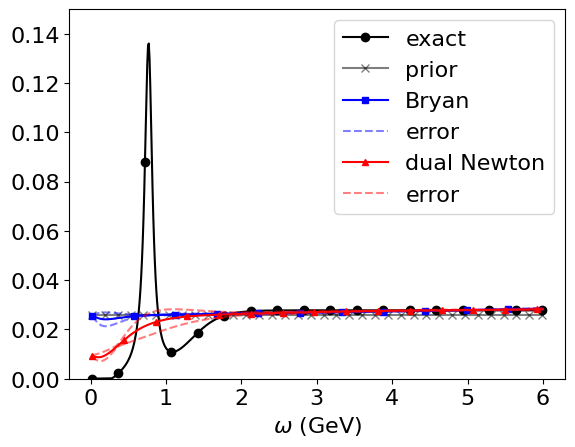}
    \caption{Plots of the parameterized $\rho$-meson spectral function \eqref{eq:rhomesonSPF}, the flat Bayesian prior, and the MEM estimates obtained using either Bryan's optimization algorithm (blue) or our dual Newton optimization algorithm (red). Plots keep number of correlator points constant (\textit{i.e.}, $N_\tau=30$) and vary the noise level $\sigma$. From left to right the noise increases as $\sigma^2 = 10^{-4}, 10^{-3}, 10^{-2}$. In the leftmost plot ($\sigma^2 = 10^{-4}$), the estimates and error bands are identical except near $\omega=0$. In the noiseless limit $\sigma^2 = 10^{-4}$ the different optimization algorithms produce similar curves, but when noise is large, MEM with Bryan's algorithm over-regularizes the solution.}
    \label{fig:rho-meson reconstructions}
\end{figure}

We reconstruct the solution from data with noise parameter $\sigma^2 = 10^{-3}, 10^{-2}, 10^{-1}$ with $N_\tau=30$ and find that as $\sigma^2 \rightarrow 0$ the MEM implemented with our dual Newton algorithm and the MEM implemented with Bryan's algorithm recover the same solution. However, our dual Newton MEM and its error estimates are more robust to noise. Bryan's MEM behavior is particularly problematic because its error bands shrink dramatically as the noise grows. This is the exact opposite of the desired behavior. Plots are presented in Figure~\ref{fig:rho-meson reconstructions}.

\begin{figure}
    \centering
    \includegraphics[width=0.33\linewidth]{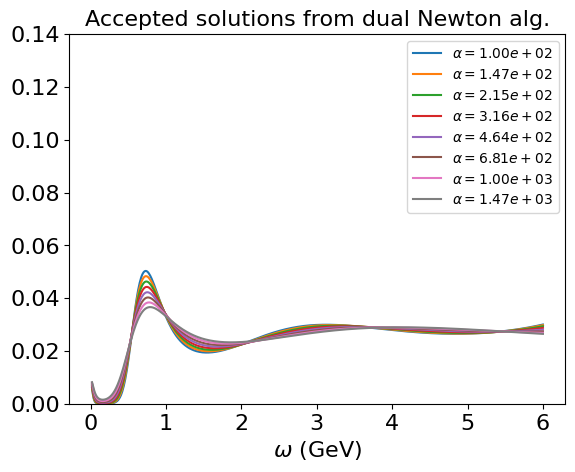}%
    \includegraphics[width=0.33\linewidth]{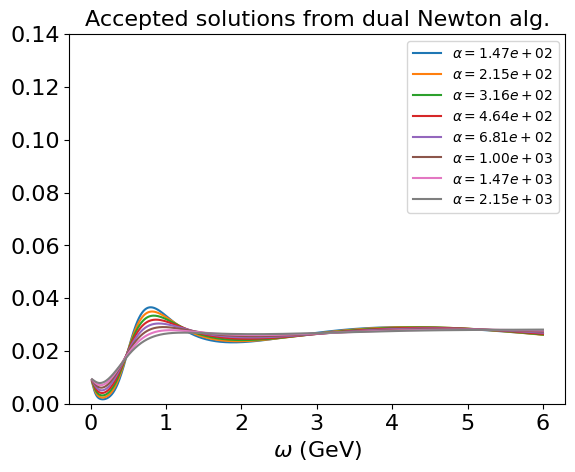}%
    \includegraphics[width=0.33\linewidth]{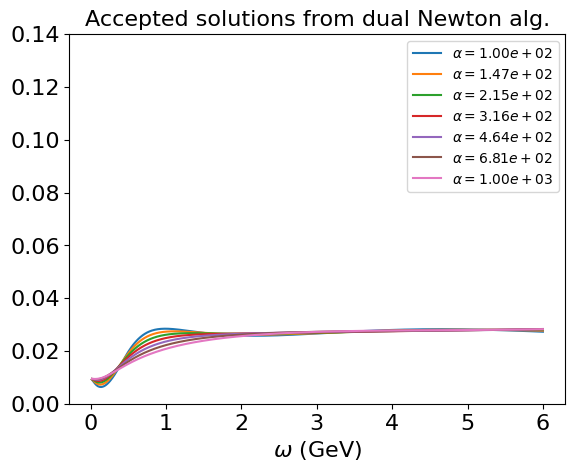}
    \includegraphics[width=0.33\linewidth]{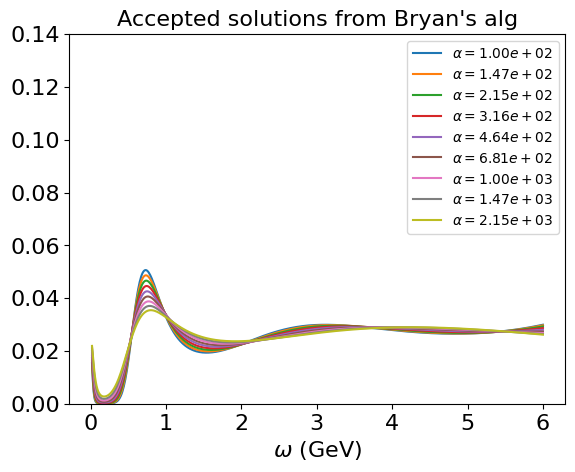}%
    \includegraphics[width=0.33\linewidth]{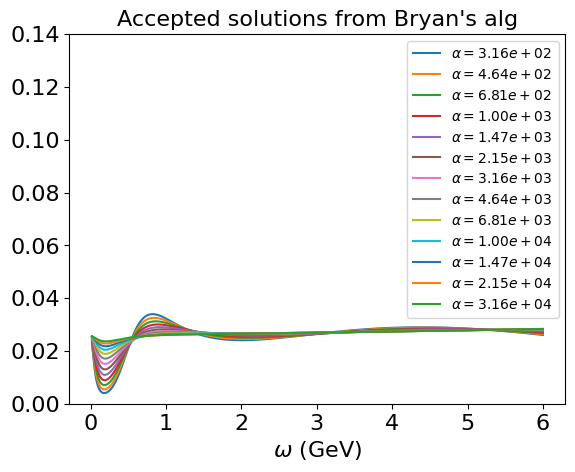}%
    \includegraphics[width=0.33\linewidth]{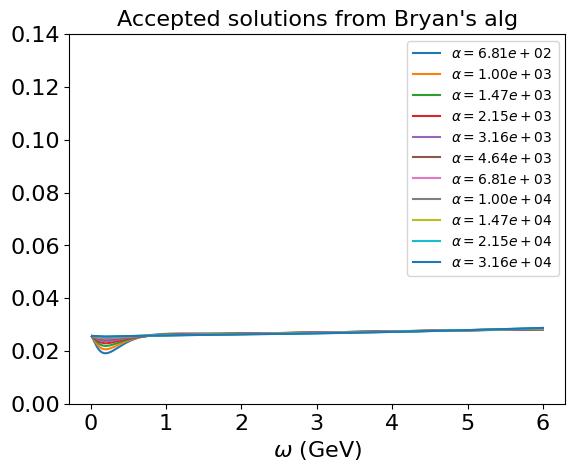}
    \includegraphics[width=0.33\linewidth]{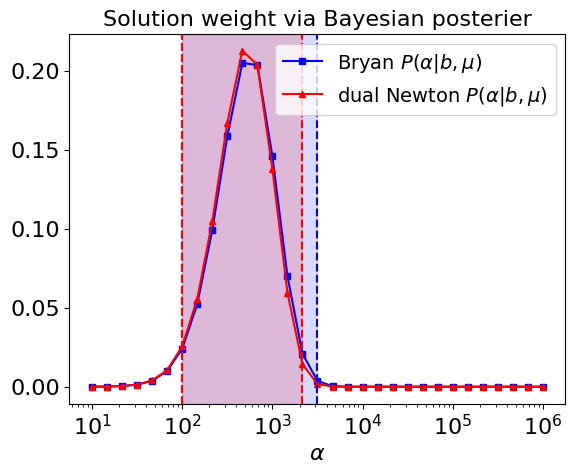}%
    \includegraphics[width=0.33\linewidth]{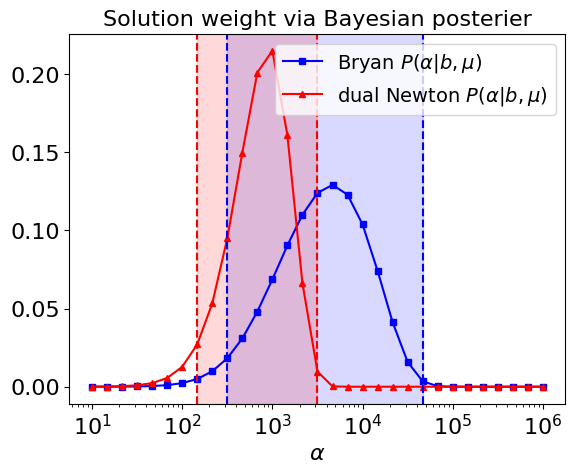}%
    \includegraphics[width=0.33\linewidth]{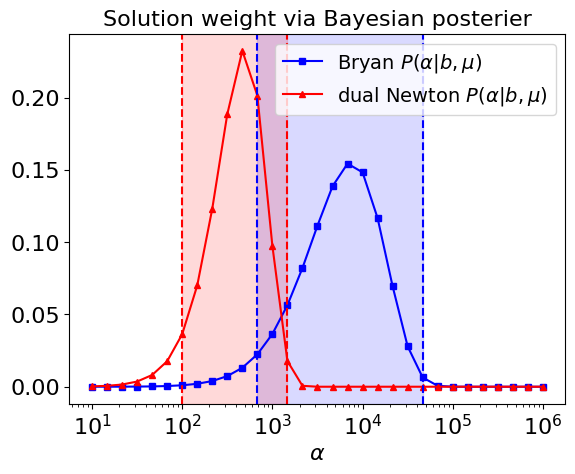}

    \caption{In rows 1 and 2 respectively, we present plots of the solutions proposed by the dual Newton algorithm and Bryan's algorithm at different regularization weights $\alpha$. All inversions use synthetic $\rho$-meson data generated from $N_\tau=30$ and the noise parameter increases from left to right, \textit{i.e.},  $\sigma^2 = 10^{-4}$ (left column),  $\sigma^2 = 10^{-3}$ (middle column), $\sigma^2 = 10^{-2}$. (right column). In the bottom row we present the posterior weighting function used to combine the various solutions. The red and blue color bands indicate the $\alpha$ domain $[\alpha_\mathbf{min},\alpha_\mathbf{max}]$ kept by the MEM with dual Newton optimizer and MEM with Bryan's optimizer respectively; see \eqref{eq:Bayesianweighting} for details. In all cases the posterior weighting function associated with the dual Newton method has a sharper peak at smaller $\alpha$ value than Bryan's primal approach. Comparing posterior weighting function across columns, as noise increases the peak of the posterior weighting function for Bryan's primal approach moves rightward. In comparison, the posterior weighting function for the dual Newton changes little.}
    \label{fig:rho-meson_proposals}
\end{figure}

To help interpret the reconstructed signals, we compare the individual $\alpha$ solutions proposed by Bryan's algorithm and by the dual Newton algorithm. We find that, given the same $\alpha$ value, the solution obtained by the dual Newton algorithm deviates further from the default model than the that obtained by Bryan's algorithm. Additionally, we compare the posterior weighting functions and find that the MEM with dual Newton optimization selects solutions which weight the data more (\textit{i.e.}, favoring solutions from lower values of $\alpha$). Both the proposed solutions and the posterior weighting functions are plotted in Figure~\ref{fig:rho-meson_proposals}. 

\begin{figure}
    \centering
    \includegraphics[width=0.5\linewidth]{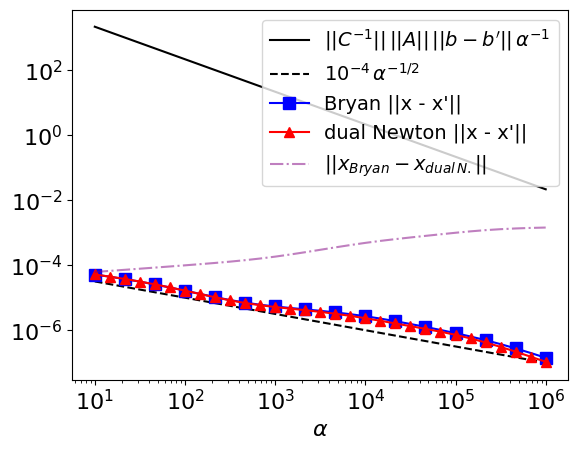}
    \caption{We study the theoretical error estimates presented in \eqref{eq:error-x} for the $\rho$-meson problem. We present the upper bound on the error in solid black; $b, b'$ correspond to noisy ($\sigma^2 = 10^{-3}$) and noiseless data. Since $\lVert C^{-1} \rVert$, $\lVert A \rVert$, and $\lVert b -b' \rVert$ are constants, the solid black line shows a $1/\alpha$ scaling. The marked red and blue curves indicate the deviation between the algorithm's solution constructed from noisy data $x$ and the solution constructed from noiseless data $x'$; these two curves are almost identical. Lastly, we plot the deviation between Bryan's solution and the dual Newton's solution from noisy data, indicated by $\lVert x_{Bryan} - x_{dual \, N.}\rVert$, which shows that while the error scaling may be similar, the constructed solutions differ.}
    \label{fig:rho-meson_errorbounds}
\end{figure}

For the discretization discussed above, $\lVert A \rVert \approx 4 \times 10^2$, $\lVert C^{-1} \rVert \approx 1 \times 10^6$, and for all $\sigma^2 = 10^{-4}, 10^{-3}, 10^{-2}$ the Euclidean norm of $b$ is $\lVert b / b_0 \rVert \approx 1 $. We numerically investigate the error bound given in \eqref{eq:error-x} and find that both Bryan's algorithm and the dual newton algorithm scale better than the theoretical bounds and that the scaling is accompanied by a small coefficient $\approx 10^{-4}$. Furthermore, we find that solutions produced by Bryan's algorithm's show greater resemblance to the default model $\mu$ for smaller values of $\alpha$ than solution's produced by the dual Newton algorithm. This manifests as deviations in the solutions at larger $\alpha$ values. These results are plotted in Figure \ref{fig:rho-meson_errorbounds}.

\subsection{Warm Dense Matter Test Problems}
We consider the extraction of dynamic structure factors (DSF) from warm dense matter (WDM) uniform electron gas (UEG) PIMC simulations. The peak in a DSF indicates a quasi-particle excited state known as a plasmon. The peak broadens with increasing temperature and tightens with increasing density. In application, having a model of the DSF allows x-ray Thomson scattering diagnostics at free electron laser facilities to estimate both the density and temperature of their plasmas. For more details see the review of recent progress given by Bonitz et al. \cite{bonitz2020ab}.

\subsubsection{Synthetic Correlation Data from UEG Dynamic Structure Factor}
The uniform electron gas (UEG)~\cite{quantum_theory,dornheim2018UEG,loos} is one of the most well-studied systems in statistical physics and quantum chemistry, making it very suitable for assessing an algorithm's performance and uncertainty quantification. A DSF can be produced from a susceptibility $\chi$ via the fluctuation dissipation theorem~\cite{quantum_theory}, in atomic units, as
\begin{align} \label{def_FDT}
    S(k,\omega) = - \frac{2}{n} \frac{\text{Im} \chi(k,\omega)}{1-e^{- \beta \hbar \omega} },
\end{align}
where the susceptibility $\chi$ quantifies the gas's density response to an external potential. 
The simplest susceptibility model satisfying the frequency sum rule and screening sum rule is Lindhard's model of the non-interacting Fermi gas with finite temperature corrections $\chi^{0}_\mathrm{Ideal}(k, \omega)$.
\begin{align} \label{eq:idealgas}
    \chi^{0}_\mathrm{Ideal}(k,\omega) = 2 \int \frac{d^3k}{(2\pi)^3} \frac{f_0 (k + q) - f_0(k)}{ (\epsilon_{k+q} - \epsilon_k) - \hbar \omega},
\end{align}
where $\epsilon_k = \hbar^2 k^2 / 2 m_e$.
We used Giuliani and Vignale's introductory textbook on the electron liquid \cite{quantum_theory} and Tolias et al.'s paper on non-interacting finite temperature electron gas \cite{tolias2024density} to implement this function.
The more sophisticated completed Mermin (CM) model also satisfies the frequency sum rule and screening sum rule and includes collisions that respect number and momentum \cite{chuna2024conservative}. For the single species electron gas the CM model is given
\begin{subequations}
\label{eq:CMsusceptibility}
\begin{align}
    \chi^0_{CM}(k,\omega + i \nu) = \cfrac{\chi^0_\mathrm{Ideal}(k, \omega + i \nu)}{ 1 - \cfrac{i}{\omega_{\tau} \tau} \left( 1 - \cfrac{\chi^0_\mathrm{Ideal}(k, \omega + i \nu)}{\chi^0_\mathrm{Ideal}(k, 0)} \right)  - \frac{i m \omega}{ k^2 n_0 \tau} \chi^0_\mathrm{Ideal}(k, \omega + i \nu) },
\end{align}
\end{subequations}
where $\nu$ is the collision frequency. For both $\chi^{0}_\mathrm{Ideal}$ and $\chi^{0}_\mathrm{CM}$ the mean field interactions are included as \cite{BohmPinesRPA1951, BohmPinesRPA1952, BohmPinesRPA1953}
\begin{align} \label{eq:RPAcorrection}
    \chi(k,\omega + i \nu) = \frac{\chi^0(k,\omega + i \nu)}{1 - v(k) \chi^0(k,\omega + i \nu)},
\end{align}
where $\nu=0$ for the ideal susceptibility. We take the particle-particle interaction $v(k)$ as a screened coulomb $v(k) = 4 \pi e^2 / (k^2 + \kappa^2)$. For this uniform electron gas, we use the un-screened limit (\textit{i.e.}, $\kappa=0$).
  
In this test problem, we generate synthetic samples by multiplying the Laplace kernel with the mean field corrected completed Mermin DSF model. We consider the UEG at density $n_e = 1.611 \times 10^{21} \, \text{cc}^{-1}$ which corresponds to a Wigner-seitz radius $r_s = 10$ Bohr. The temperature is $ k_B T = 0.5$ eV, equivalently $\beta =  1/ k_B T  =54.301$ Hartree, which corresponds to a quantum degeneracy parameter of $\Theta = k_B T / E_F = 1$; these conditions match the conditions of the PIMC UEG simulation, whose correlation estimates will be considered next. \edit{For this $n$ and $T$, the plasma frequency is $\omega_{p,e} = \sqrt{4 \pi n e^2 / m_e } = 0.05476$ Hartree and Debye wavenumber is $k_{D,e}= \sqrt{4 \pi n e^2 / T } =0.40362 \, \text{Bohr}^{-1}$}{For this system the electron plasma frequency is $\omega_{p,e} = \sqrt{3 / r_s^3} = 0.05476 \, \text{Hartree}$ and Fermi wavenumber is $k_{F,e}= (9 \pi / 4 )^{1/3} / r_s = 0.1919 \, \text{Bohr}^{-1}$}. The collision frequency $\nu$ is set equal to the electron plasmon frequency $\omega_{p,e}$. We discretize our frequency as $\omega_j = j \Delta \omega$ where $j \in \{1,\ldots,N_\omega\}$, $N_\omega = 1250$, $\Delta \omega = 0.0011$ Hartree. Additionally, we discretize our imaginary time as $\tau_i = i \Delta \tau$ where $i \in \{0,\ldots,N_\tau-1\}$, $N_\tau = 201$, $\Delta \tau = 0.271505$ Hartree. We introduce noise different from \eqref{eq:syntheticnoise}; we chose a standard deviation of
\begin{align}\label{eq:syntheticnoise_2}
    \sigma_i = \sigma \, b_i.
\end{align} 
This noise does not grow proportionally to $\tau$ and is more reflective of our authentic PIMC data. We generate $N_s=1000$ samples of $F(\tau)$ and from them estimate the data $b$ and the error matrix $C_{i,j} = \delta_{i,j} \sigma_i^2 / \sqrt{N_s}$. We conduct the inversions using the ideal gas with mean field correction (\textit{i.e.}, RPA) as the default model $\mu$. We follow the same Bayesian posterior weighting procedure described for \eqref{eq:Bayesianweighting} to average our collection of solutions into an estimate.

\begin{figure}[h]
    \centering
    \includegraphics[width=.33\textwidth]{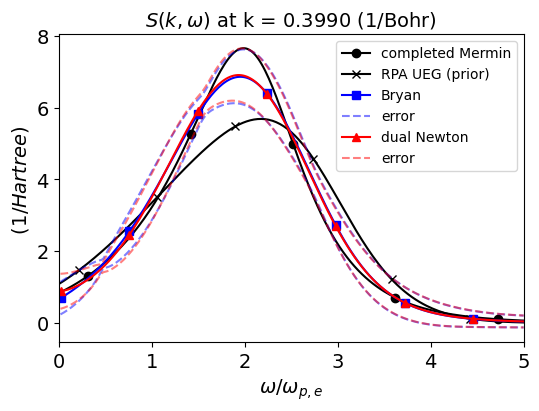}%
    \includegraphics[width=.33\textwidth]{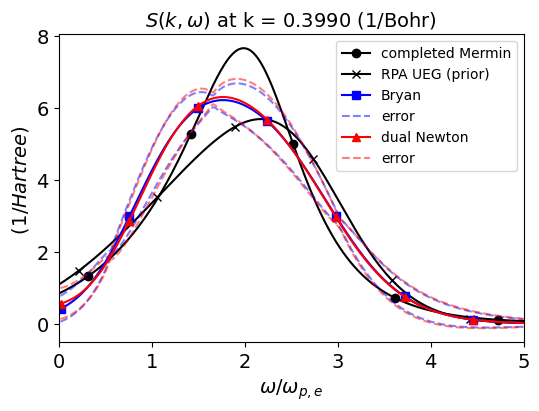}%
    \includegraphics[width=.33\textwidth]{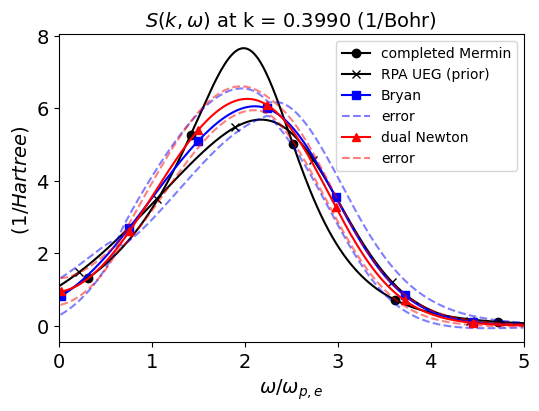}
    \includegraphics[width=0.33\linewidth]{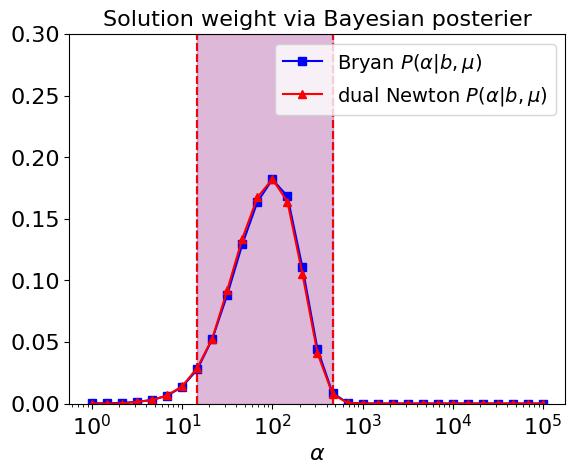}%
    \includegraphics[width=0.33\linewidth]{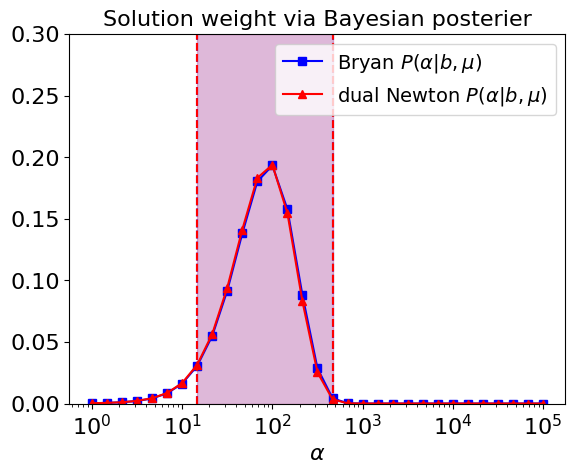}%
    \includegraphics[width=0.33\linewidth]{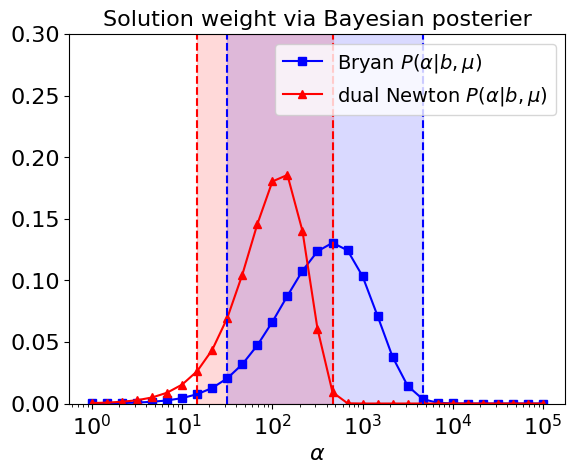}
    \caption{Plots of the estimated dynamic structure factor for the synthetic UEG example with associated posterior weighting function. From left to right, the noise level $\sigma$ increases as $\sigma = 10^{-2}, 10^{-1}, 10^{0}$. The top row contains the plots of the DSF. The completed Mermin (CM) model is the the input spectrum and desired recovery, defined in \eqref{eq:CMsusceptibility} with mean field correction \eqref{eq:RPAcorrection} included. The RPA UEG is the Bayesian prior, defined in \eqref{eq:idealgas} with mean field correction \eqref{eq:RPAcorrection} included. The MEM estimates are in either red or blue. On the bottom row the Bayesian posterior is plotted as a function of $\alpha$. We logarithmically sample alpha in $\alpha \in [10^{0}-10^5]$. The red and blue vertical color bands indicates the $\alpha$ domain is selected for either MEM averaging procedure; see \eqref{eq:Bayesianweighting} for details on how the $\alpha$ domain was selected. Comparing posterior weighting functions, as noise increases the peak of the posterior weighting function for Bryan's primal approach moves rightward. In comparison, the posterior weighting function for the dual Newton changes little.}
    \label{fig:syntheticUEG_reconstruction-curves}
\end{figure}

We first consider the DSF estimate for a small wavenumber $k = 0.3990 \, \text{Bohr}^{-1}$ and noise levels $\sigma = 10^{-2},  10^{-1},  10^{0}$ (see Figure~\ref{fig:syntheticUEG_reconstruction-curves}). In the limit of small noise, both the MEM with Bryan's algorithm and the MEM with dual Newton converge to the desired completed Mermin $S(k,\omega)$. However, as noise gets large, we see again that Bryan's algorithm breaks down, over-regularizing the solution by selecting weighting the larger $\alpha$ values. This is the same behavior seen in for $\rho$-meson investigation. 

\subsubsection{Authentic Correlation Data from UEG PIMC simulation}
The authentic WDM UEG PIMC correlation data we study arises from Dornheim and Vorberger's work \cite{Dornheim2020PIMCUEG}. They employed path integral Monte Carlo (PIMC) methods to produce Euclidean time correlation functions of the uniform electron gas at WDM conditions ($T=0.5$eV and $r_s=10$). This PIMC UEG data was analyzed by Dornheim et al. \cite{dornheim2018stochasticsamplingalg}, who used a stochastic sampling algorithm to infer the DSF. Their stochastic sampling algorithm is a forward problem approach that relies on sampling values for a parameterized form of the susceptibility; the parameterized form is determined via sum rules and additional constraints. By comparison, our inverse method has no similar guarantee of satisfying sum rules, but will work in systems where a parameterized form of the susceptibility is not known and hence has a broader range of applicability. In this section, we include the stochastic sampling results in the slot where the completed Mermin was in synthetic results because the general shape of the DSF is physical even if the stochastic estimate is not the true solution.

\begin{figure}
    \centering
    \includegraphics[width=.33\textwidth]{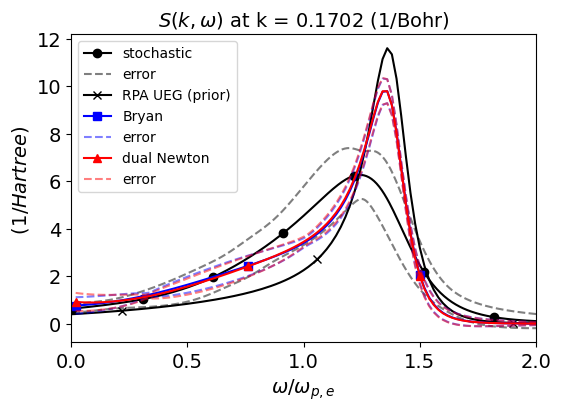}%
    \includegraphics[width=.33\textwidth]{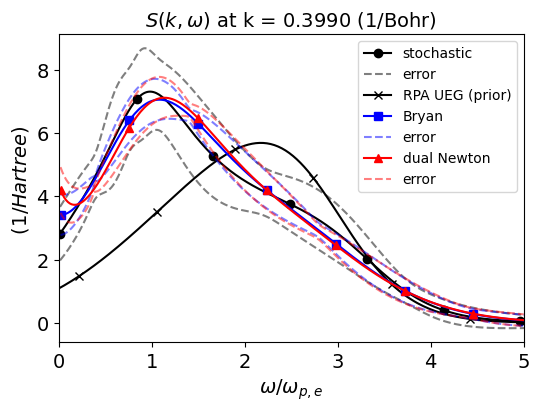}%
    \includegraphics[width=.33\textwidth]{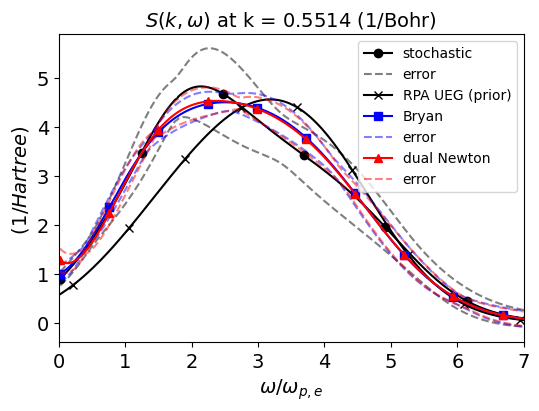}
    \includegraphics[width=.33\textwidth]{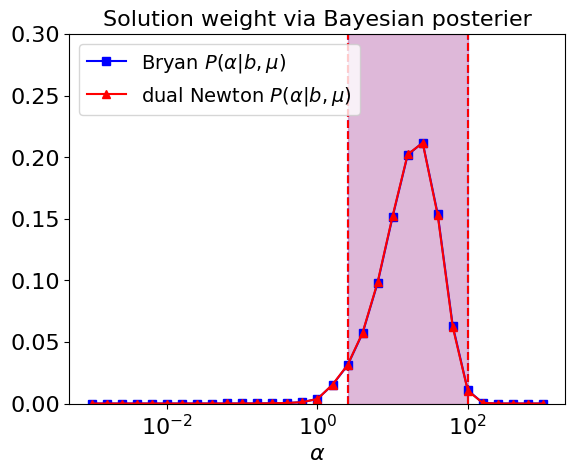}%
    \includegraphics[width=.33\textwidth]{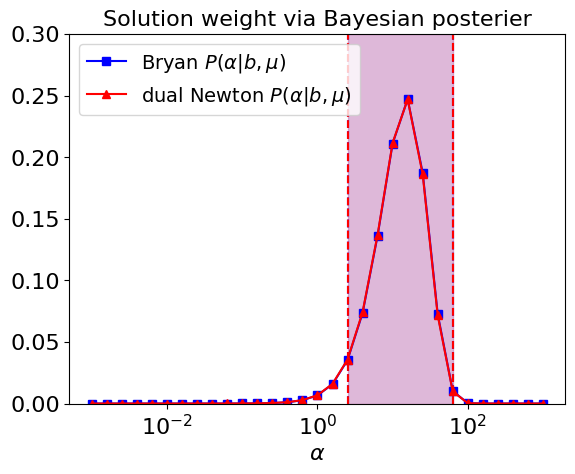}%
    \includegraphics[width=.33\textwidth]{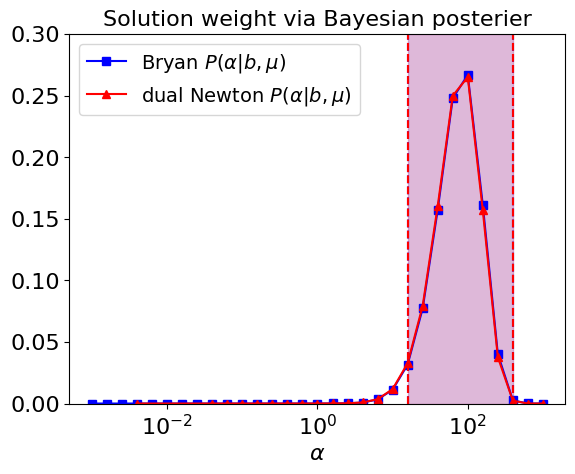}  
    \caption{Top: From left to right we vary the wavenumber $k$ considered in our DSF, for all $k$ values, the number of correlator points is fixed at $201$ and the relative noise is $\approx 10^{-3}$. The curve labeled ``stochastic'' is Tobias et al.'s estimate \cite{dornheim2018stochasticsamplingalg}, not the exact solution. The RPA UEG is the Bayesian prior, defined in \eqref{eq:idealgas}, with the mean-field correction~\eqref{eq:RPAcorrection} included. Bottom: Plot of the posterior weighting function associated with the above estimates. We collect solutions for $\alpha \in [10^{-3}-10^3]$. The red and blue color bands indicate the $\alpha$ domain kept by the MEM with dual Newton optimizer and MEM with Bryan's optimizer respectively; see \eqref{eq:Bayesianweighting} for details on how the $\alpha$ domain was selected. We find strong overlap between the Bryan and dual Newton approach is readily explained by the small relative noise in the data.}
    \label{fig:PIMCUEG_reconstruction}
\end{figure}

We present plots of the reconstructed $S(k,\omega)$ at a small, middle and large wavenumber $k = 0.1702$, $0.3990$, $0.5513$ $\text{Bohr}^{-1}$. These plots show excellent agreement with the stochastic method at large $k$, but poorer agreement at small $k$. Anecdotally, we found that if the Bayesian prior is sharply peaked, then the solutions deviate less from the prior. Furthermore, both MEM approaches produce solutions that have less ringing than the stochastic model; ringing at large $\omega$ is a common concern for the BRM entropic approach \cite{Fischer2018smoothBR}. Plots are given in Figure \ref{fig:PIMCUEG_reconstruction} top row.

We also present the normalized Bayesian posterior used to combine the $\alpha$-solutions into an estimate. Since the error in the correlation function is small, approximately $10^{-3}$, we do not expect that Bryan's Bayesian posterior will differ from the dual Newton's Bayesian posterior. This is reflected in Figure \ref{fig:PIMCUEG_reconstruction} bottom row \edit{}{and explains why the MEM with Bryan and the MEM with dual Newton estimates are indistinguishable from the PIMC data.}.  

\begin{figure}
    \centering
    \includegraphics[width=.33\textwidth]{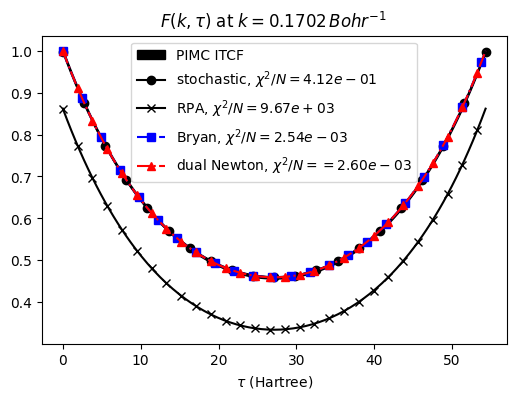}%
    \includegraphics[width=.33\textwidth]{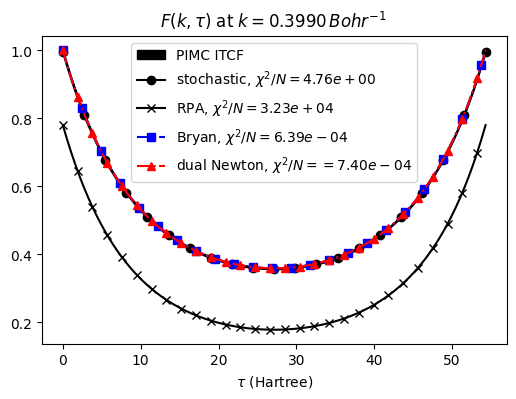}%
    \includegraphics[width=.33\textwidth]{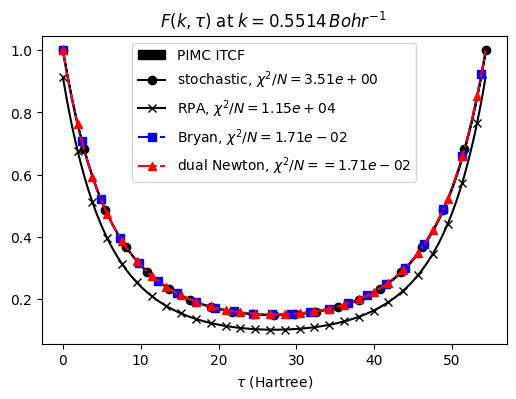}
    \caption{Plots of the periodic Laplace transformed DSF estimates presented in the top row of Figure~\ref{fig:PIMCUEG_reconstruction} (transform defined in ~\eqref{eq:acproblem-periodic}). We include the imaginary time correlation function (ITCF) data obtained via PIMC UEG data for comparison. The periodic Laplace transform of the MEM DSF estimates are visually indistinguishable from the observed ITCF data, while the periodic Laplace transform RPA DSF shows clear differences. This indicates that the solution is not so heavily regularized that the entropic prior dominates the cost function.}
    \label{fig:PIMCUEG_correlation}
\end{figure}

We plot the correlation function that would be produced from the reconstructed DSFs and compare to the PIMC UEG data. This helps us compare to the stochastic sampling method which accepted and averaged over DSF's based on the proposed DSF's ability to ``reproduce the [correlation function] within the Monte Carlo error bars''. We see that both MEM with the dual Newton and Bryan's optimization produce correlation functions that are indistinguishable from the data. The associated reduced chi-squared value $\chi^2/N_\tau = (Ax-b) C^{-1} (Ax-b) / N_\tau$ reflects this. The chi-square values also inform us that the MEM estimates match the data more closely than the stochastic procedure. Plots of the correlation functions are given in Figure \ref{fig:PIMCUEG_correlation}.

\begin{figure}
    \centering
    \includegraphics[width=\textwidth]{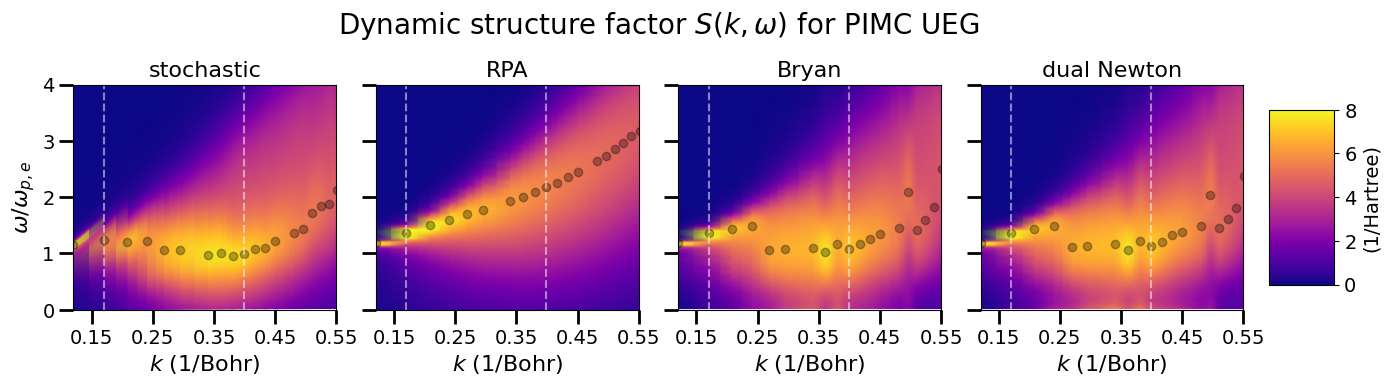}%
    \caption{Heatmaps of the dynamic structure factor $S(k,\omega)$. The $S(k,\omega)$ estimate from \cite{Dornheim2020PIMCUEG} is labeled stochastic sampling. The Bayesian prior, indicated RPA is defined in \eqref{eq:idealgas} with mean field correction \eqref{eq:RPAcorrection} included. The estimates obtained by MEM with Bryan's algorithm and the estimates obtained by MEM with the dual Newton optimization are also included. Dashed white lines have been placed at $k = 0.170153$, $0.399044$, and $0.551358 \, \text{Bohr}^{-1}$. These lines indicate the $k$ values plotted in Figures~\ref{fig:PIMCUEG_reconstruction} and \ref{fig:PIMCUEG_correlation}. For each $k$ value the location of the DSF peak is indicated with a black dot; together they indicate the dispersion relation. We see that the MEM DSF estimates have a qualitatively different dispersion relation than the RPA Bayesian priod, which better resembles the stochastic estimate.}
    \label{fig:PIMCUEG_Skw}
\end{figure}

We also investigate the full DSF $S(k,\omega)$ estimate and find that both Bryan's algorithm and the dual Newton algorithm produces DSF's that contain a dip in the dispersion relation occurring for $k \in [0.25, 0.45] \,\text{Bohr}^{-1}$, which is known as a roton feature~\cite{Dornheim_Roton_2022}. Comparing to the stochastic approach the MEM approaches do not produce as smooth of a dispersion relation. We interpret this as a result of the data indicating there is a roton feature, but the Bayesian prior (\textit{i.e.}, RPA) not having this feature and then some compromise is reached. This serves as a reminder that all forms of entropic regularization produce a unique solution that is biased towards the Bayesian prior. Two possible ways to address this are using improved default models such as the \emph{static approximation} in the case of the warm dense UEG~\cite{Dornheim_PRL_2020_ESA} or to conduct the entire $S(k,\omega)$ inversion at once rather than divided into individual $k$ inversions. Plots of the full DSF are given in Figure~\ref{fig:PIMCUEG_Skw}. 

\section{Conclusions \label{sec_conclusions}}
In this work we introduce the dual formulation of the entropy regularized least squares cost function and analytic error bounds on the dual estimate. The advantage of the dual formulation is that optimizing the dual produces the same solution as optimizing primal problem, but the dual cost function has guaranteed strong convexity, differentiability, and lower dimension. This makes the dual problem robust to noise, tractable to second order optimization methods, and computationally cheaper. Of particular importance is that the dual optimization problem is formulated in the kernel's row space, and thus the resolution of the spectral function does not impact the dimension of the dual optimization. In short, dual formulation has clear advantages for the analytic continuation problem, where the kernel is ill-conditioned and the spectral function is continuous. The error bounds establish that the dual optimization produces corresponding primal solutions that converge to the desired minimum. Furthermore, these error estimates provide insight into how a solution depends on the noise and on the regularization parameter $\alpha$. Numeric tests showed that the theoretical upper bound on the error was many orders of magnitude larger than our actual error. 

Many advantages of our dual approach are particular to the problem formulation that has been given in Section~\ref{subsec_ProblemFormulation}. These advantages do not carry over to all of the regularization terms which are currently being used in the literature (\textit{e.g.}, Lasso L1 regularization \cite{Otsuki_PRE_2017}). Further, in the rare case where a closed form of the solution is known, iterative optimizations in either a dual formulation or a primal formulation will not outperform the closed solution (\textit{e.g.} ridge L2 regularization, the Backus-Gilbert method \cite{backus1968, backus1970}, and its smooth variant \cite{hansen2019extraction}). The analytic continuation community needs to consider whether their problem formulation may benefit from a dual formulation.

We have compared our dual Newton algorithm with Bryan's heavily cited algorithm\edit{, which does not optimize the entropy regularized least squares cost function, but only an approximate version}. \edit{From these comparisons, we conclude that there are two advantages}{In section~\ref{sec_methods}, we discuss two theoretical advantages} of our approach. First, unlike Bryan's algorithms we search for the best positive normalization of the estimate\edit{ rather than force the estimate to match the default model's positive normalization}{}. We argue that this capability is important for the periodic Laplace transform (\textit{i.e.}, finite temperature systems). Second, our approach optimizes in a reduced search space without sacrificing singular basis vectors. This addresses Rothkopf's results which showed that removing singular vectors prevents the MEM from representing sharply peaked functions at large $\omega$ \cite{rothkopf2020bryan}. In section~\ref{sec_results}, we demonstrate that when the noise is small both algorithms yield the same mean and variance estimates, but in presence of large noise our algorithm differs from Bryan's algorithm. Knowing that Bryan's algorithm makes an assumption that the gradient of the fidelity term lies in the singular space while our algorithm does not, these differences suggest that noise also violates Bryan's assumption. In summary, Bryan's algorithm relies on assumptions that the noise is small and the desired function is sharply peaked at small $\omega$. As a result we echo Rothkopf's statement, ``either systematically extend the search space within the MEM or abandon the MEM in favor of one of the many modern Bayesian approaches developed over the past two decades.'' We have expanded the search space within MEM and in doing so we have not seen the introduction of prominent ringing artifacts. 

Future applications of this method are broad; the dual MEM routine used in this work can handle a generic kernel, so it can be readily applied to the applications mentioned in the introduction section. For the plasma physics community, future work physics-oriented work could apply this method to other number densities and temperatures to observe how the roton feature varies with these parameters. Future methods-oriented work could extend the cost function and solve for all of $S(k,\omega)$ simultaneously, rather than solving each $k$ individually. The same sentiment is shared by the Lattice QCD community, which wants to estimate spectral functions $\rho(\omega, T)$ for all temperatures simultaneously rather than a given $T$. In both cases, solving for all values simultaneously would enhance smoothness across the additional variable. Additionally, for the plasma physics community this work has shown that entropic regularization can invert the Laplace transform, which matches sentiments of other fields and opens the possibility of conducting instrument deconvolution using the Laplace transform instead of the Fourier transform. Lastly, future work may wish to improve over the Newton-conjugate gradient method discussed in \ref{app:newton-cg} or explore alternatives to using the discovered positive normalization of $x$ for the default model $\mu$. 

\appendix
\section{Description of the dual algorithm \label{app:newton-cg}}

This section outlines the Newton-Krylov method~\cite{kelleySolvingNonlinearEquations2003,knollJacobianfreeNewtonKrylov2004} used to solve the MEM problem presented in~\eqref{eq:primal}. 
The purpose of this method is to obtain the scaling $\bar Z$ that minimizes the value function $V(Z)$ defined by~\eqref{eq:scaledprimal} and to obtain a corresponding primal solution $\bar x$ of~\eqref{eq:primal}.


Duality theory establishes a direct relationship between the primal and dual solutions through smooth mappings, as given by~\eqref{eq:primal-dual-maps}. Solving the dual problem yields the same optimal value as the primal problem; however, because the dual cost function is strongly convex and differentiable, it is amenable to a simpler optimization process.

The underlying duality theory that gives rise to the smooth primal-dual maps implies that
\[
V(Z) = \min\{p(x,Z\mid\mu/Z)\mid x\in Z\Delta_{N_\omega}\} = -\min\{d(y,Z\mid\mu/Z) \mid y\in\rr^{N_\tau}\}.
\]
Since the dual objective $d(y,Z\mid\mu/Z)$ is strongly convex in $y$, the dual minimizer $\bar y_Z$, which depends on the scaling $Z$, is unique and the Hessian $\nabla^2_y d(\bar y_Z,Z\mid\mu/Z)$ is positive definite and hence nonsingular. Thus, 
\[
V(Z) = -d(\bar y_Z,Z\mid\mu/Z)
\]
is differentiable. Applying the chain rule, one deduces that its derivative with respect to $Z$ is
\begin{equation}\label{eq:value-function-deriv}
V'(Z) = -\logexp(A^*\bar y_Z\mid\mu) + \log Z + 1.
\end{equation}
Observe that the second derivative $V''(Z) = 1/Z$ is positive for all positive scalings $Z$, and thus $V$ is strictly convex and achieves its minimum value when
\begin{equation}\label{eq:v-prime-zero}
V'(Z)=0.
\end{equation}
Minimizing $V(Z)$ over $Z$, one obtains an optimal pair ($\bar y, \bar Z)$ and consequently, the corresponding primal solution $x(\bar y)$.

The optimization over $Z$ is conducted using a binary search to obtain a root of the equation \eqref{eq:v-prime-zero}, which corresponds to the optimal $Z$. Each iteration of the binary search algorithm requires solving the dual problem \eqref{eq:dual} using a trust-region Newton-Krylov method implemented by the Julia package \texttt{JSOSolvers.jl}~\cite{Migot_JSOSolvers_jl_JuliaSmoothOptimizers_optimization_2023}. In practice, the dual problem becomes easier to solve at each iteration because the current dual solution can be used to warm start the subsequent dual solve. The linear Krylov solver requires only matrix-vector products with the Hessian of \eqref{eq:dual} (\textit{i.e.}, $\nabla^2_y d(\bar y_Z,Z\mid\mu/Z)$). The overall computational cost of the resulting algorithm is dominated by products with the operator $A$ and its adjoint $A^*$.

For this class of problems, the Newton-Krylov method is guaranteed to produce a sequence of dual iterates that converge to the unique solution. Locally, Newton's method is expected to converge quadratically, although inexact solves of the Newton equation may cause the method to converge only superlinearly. Additionally, this computation generates a corresponding primal sequence $\{x(y^k)\}$, defined by the smooth map~\eqref{eq:smooth-map-x}, which converges to the unique global solution $x(y^*)\in\Delta_n$, as required. We terminate the algorithm when the gradient
\begin{equation}\label{eq:nonlinear-equation}
F(y,Z) = \begin{bmatrix}
    \nabla_y d(y,Z\mid\mu)\\
    V'(Z)
\end{bmatrix}
\end{equation}
drops below absolute and relative tolerances, $\epsilon_{a}$ and $\epsilon_{r}$, as $\lVert F(y,Z) \rVert \le \epsilon_{a} + \epsilon_{r}(1+\|b\|)$.
Our experiments all use $\epsilon_a = \epsilon_r = 10^{-6}$.

\section*{Acknowledgments}
This work was partially supported by the Center for Advanced Systems Understanding (CASUS) which is financed by Germany’s Federal Ministry of Education and Research (BMBF) and by the Saxon state government out of the State budget approved by the Saxon State Parliament. This work has received funding from the European Union's Just Transition Fund (JTF) within the project \emph{R\"ontgenlaser-Optimierung der Laserfusion} (ROLF), contract number 5086999001, co-financed by the Saxon state government out of the State budget approved by the Saxon State Parliament.
This work has received funding from the European Research Council (ERC) under the European Union’s Horizon 2022 research and innovation programme (Grant agreement No. 101076233, "PREXTREME"). 
Views and opinions expressed are however those of the authors only and do not necessarily reflect those of the European Union or the European Research Council Executive Agency. Neither the European Union nor the granting authority can be held responsible for them.
The PIMC calculations were partly carried out at the Norddeutscher Verbund f\"ur Hoch- und H\"ochstleistungsrechnen (HLRN) under grant mvp00024, and on a Bull Cluster at the Center for Information Services and High Performance Computing (ZIH) at Technische Universit\"at Dresden. M.P. Friedlander is partially supported by a grant from the Natural Science and Engineering Council of Canada. 

\bibliographystyle{unsrt}
\bibliography{references.bib}

\end{document}